# Global Vulnerability Assessment of Mobile Telecommunications Infrastructure to Climate Hazards using Crowdsourced Open Data

Edward J. Oughton[1,2*], Tom Russell[2], Jeongjin Oh[3], Sara Ballan[3], and Jim W. Hall[2]

[1]College of Science, George Mason University, Fairfax, VA, USA

[2]Environmental Change Institute, University of Oxford, Oxford, Oxfordshire, UK

[3]World Bank, Washington DC, USA

*Corresponding author: Edward J. Oughton (e-mail: eoughton@gmu.edu)

Address: College of Science, George Mason University, 4400 University Drive, Fairfax, VA

## Abstract

The ongoing change in Earth`s climate is causing an increase in the frequency and severity of climate-related hazards, for example, from coastal flooding, riverine flooding, and tropical cyclones. There is currently an urgent need to quantify the potential impacts of these events on infrastructure and users, especially for hitherto neglected infrastructure sectors, such as telecommunications, particularly given our increasing dependence on digital technologies. In this analysis a global assessment is undertaken, quantifying the number of mobile cells vulnerable to climate hazards using open crowdsourced data equating to 7.6 million 2G, 3G, 4G and 5G assets. For a 0.01% annual probability event under a high emissions scenario (RCP8.5), the number of affected cells is estimated at 2.26 million for tropical cyclones, equating to USD 1.01 billion in direct damage (an increase against the historical baseline of 14% and 44%, respectively). Equally, for coastal flooding the number of potentially affected cells for an event with a 0.01% annual probability under RCP8.5 is 109.9 thousand, equating to direct damage costs of USD 2.69 billion (an increase against the baseline of 70% and 78%, respectively). The findings demonstrate the need for risk analysts to include mobile communications (and telecommunications more broadly) in future critical national infrastructure assessments. Indeed, this paper contributes a proven assessment methodology to the literature for use in future research for assessing this critical infrastructure sector.

## 1. Introduction

Climate change is one of the major challenges of our time, with society and the economy needing to rapidly adapt to an increase in the frequency and severity of disasters (Wuebbles et al., 2017; Pörtner et al., 2022). Adaptation of critical infrastructure is therefore essential particularly for 'lifeline utilities', such as energy, transportation, water, waste, and telecom networks. However, infrastructure assessment research focusing on climate vulnerability has largely neglected telecommunication infrastructure, defined here as consisting of 2G, 3G, 4G and 5G cells, local fixed broadband networks, long-distance land and submarine fiber networks, satellite networks, and cloud data centers.

For example, whereas there have been comprehensive evaluations for other critical infrastructure sectors such as transportation (Koks et al., 2019), the global vulnerability assessment of telecommunication infrastructure from natural disasters has received little systematic focus in the literature. Very often there has been a focus on field-based post-disaster damage evaluation, after flooding or tropical cyclones. Examples of telecommunication infrastructure damage assessments for natural disasters between 2005-2011 include the Katrina, Gustav and Ike hurricanes, Chile's 2011 earthquake and tsunami, New Zealand's 2011 earthquake in Christchurch, and the Great Earthquake and Tsunami in the Tohoku Region of Japan (Krishnamurthy et al., 2016; Kwasinski, 2011). Unfortunately, more recent cost-benefit assessments carried out by the World Bank, focusing on strengthening infrastructure assets, telecommunication infrastructure has been excluded (Hallegatte et al., 2019).

The contribution of this analysis is to begin to fill this literature gap by developing new methods to evaluate the climate hazard vulnerability of mobile cellular infrastructure. This is important because quantitative scenario analysis can produce geospatial hazard information that can be used in multiple different ways. Firstly, such information can help support adaptation decisions by policy makers and/or Mobile Network Operators, with regards to strengthening existing infrastructure assets (Hall et al., 2017). Secondly, such information can inform location planning of future assets, along with building design codes for extra protection in areas of higher vulnerability from multi-hazard threats (Pourghasemi et al., 2020; Yousefi et al., 2020). Finally, hazard information can enable risk exposure to be shared among many different parties, for example, by businesses and households sharing risk with insurers (who in turn share risk with re-insurers) (Broberg, 2020; Dallimer et al., 2020). To move towards these adaptation options, it is imperative that a strong vulnerability assessment method is identified, particularly as such an approach can support the inclusion of this sector in future cross-sectoral appraisals.

Given the issues raised in this introduction, the following research questions are articulated to guide the climate vulnerability assessment of mobile cellular infrastructure:

    I.    How many cellular assets are potentially vulnerable to climate hazards, here comprising coastal flooding, riverine flooding, and tropical cyclones?

    II.    What magnitude of economic damage may occur from different climate hazard scenarios, for example from flooding, riverine flooding, and tropical cyclones?

In the following section (Section 2) a literature review is presented to explore the range of related studies assessing climate hazard impacts on infrastructure. Next, a method is detailed in Section 3 to obtain the necessary results to answer the research questions articulated. Subsequently, Section 4 will report the results obtained before a discussion of the key findings is undertaken in Section 5. Finally, conclusions are given in Section 6.

## 2. Literature review

The telecommunications sector consists of a wide range of infrastructure assets, including cellular towers, copper and fiber optic cables, switches, satellites, Wi-Fi and small cell hotspots, and long-range microwave radio links (Hall et al., 2016; Oughton et al., 2018; Pant et al., 2020). There are challenges to obtaining data on many of these assets as they are very often built by private companies (and thus commercially sensitive), as well as being hard to identify from other methods (e.g., aerial or satellite imagery) (Mohebbi et al., 2020; Oughton and Mathur, 2021). Such data limitations have often constrained analysis on this sector, leaving many assessments focused on small case study areas.

Disrupted mobile communications in disaster zones leads to poor coordination and decision making, as disaster response professionals and resources are not directed to areas most in need (EL Khaled and Mcheick, 2019; Kaja et al., 2021). Moreover, when emergency organizations are unable to effectively coordinate and communicate, this can also lead to catastrophic outcomes for both human life and properties, particularly as threats from hazard environments evolve over time during disaster events (Rosario-Albert and Takahashi, 2021). Given the convenience of owning a cell phone, when a disaster does take place, this is often the only communication device available for first responders and for citizens attempting to contact others (Foster, 2011). Therefore, telecommunication sites need to be provisioned to provide resilient service in such circumstances, with decision-makers selecting the most optimal network architecture and design specifications to ensure reliable service (Gallarno et al., 2023).

There are only a very limited number of post-event damage assessments of telecommunication networks in the literature, evaluating the extent to which water and wind damage interrupt telecommunication assets, and the voice and data services they provide (Kwasinski, 2013; O'Reilly et al., 2006). Often these have been carried out for major hurricanes when evaluating impacts to a range of critical infrastructure sectors including power networks. Following Hurricane Katrina, on-site surveys of damage were conducted to observe the impacts on broadband networks, with substantial disruption taking place to centralized network elements (not just distributed network assets) (Kwasinski et al., 2009). Overall, Hurricane Katrina is estimated to have led to the failure of more than 1,000 cellular sites, although 80% were able to be put back online within one week (as base station realignment and restoring power was often sufficient), whereas the 20% more severely damaged were in Gulf coast areas more severely affected (and thus required more substantial repairs or full replacement) (Federal Communications Commission, 2006). This contrasts with 320 cell site failures during Hurricane Harvey, equating to approximately 4.1% of the 7,804 cell sites in coastal counties exposed in Texas and Louisiana (Federal Communications Commission, 2017).

Generally, two main failure effects can occur. Firstly, the impact of flooding leads to direct floodwater contact to active electronic components, including backup generators, fuel tanks, as well as network equipment, leading to disruption. During Hurricane Katrina, while the elevated mounting of some cellular assets helped reduce recovery time by avoiding flood damage in the New Orleans area, many cellular sites had a lack of uniformity in cell site construction practices. Indeed, often only part of the base station equipment present may be elevated above the flood plain, and therefore protected from floodwaters, leaving some telecom components or necessary energy equipment vulnerable. Secondly, logistical issues can cause fuel required for maintaining backup generators to never reach key assets, leading to a loss of power (Kwon et al., 2016). When this damage does occur, rebuilding costs can be very high. Following damage caused by Hurricane Matthew in The Bahamas, telecommunications constituted up to 63% of the total cost for replacing damaged infrastructure assets across the Archipelago, as many sites had been flooded and affected by wind damage (Bello et al., 2020).

The impact on cellular telecommunications infrastructure from natural hazards has been characterized by five potential damage states (Kwasinski, 2011), which include:

I.   The absence of permanent onsite backup electricity generation equipment, combined with exhausted onsite backup battery packs, leading to loss of service.

II.  Failure of permanent onsite backup electricity generation equipment leading to loss of service. This outcome could occur through usage of all backup fuel and poor maintenance of generation equipment leading to a failure to start.

III. Substantial damage to onsite backup electricity generation equipment while leaving communications equipment (e.g., active radio equipment) fully functioning. Frequently onsite backup electricity generation equipment may be placed at ground level, even if active radio electronic equipment is elevated to avoid floodwater damage.

IV.  Severe loss of functionality to both active electronic radio equipment and onsite backup electricity generation equipment.

V. Catastrophic damage, leading to failure, in other infrastructure assets necessary to provide normal service. For example, this could be physical damage to fiber optic backhaul cables or other ground-based switching equipment, or misalignment of wireless backhaul units leading to interrupted or severely degraded services.

Hazards also pose a major risk to the civil engineering structures used to mount radio antennas. Indeed, the actual tower design can have a large impact on how affected an asset may be, with three main designs including monopoles, self-supported structures, or guyed structures. Often monopoles extend up to 60 meters and consist of hollow cantilevered structures and are commonly used in urban areas where land costs are prohibitively high for other designs. For greater height, lattice structures are used to provide self-supported towers up to 120 meters. Finally, for heights up to 500 meters, guyed structures consist of a modest narrow mast with lateral stability, supported by multiple ground anchors via tensioned guy cables (Suryakumar et al., 2020). In rural areas where land costs are lower, it is more common to utilize higher guyed structures.

Cell phone towers have typically been built to withstand certain hazard exposures (e.g., wind speeds) above which they may fail. However, they are still vulnerable and at risk from hazards below these thresholds too, for example, during high wind speeds during hurricanes. During Hurricane Maria in 2018, winds of up to 280 km/h felled more than 90 percent of the cell phone towers in Puerto Rico. Risks are more moderate at lower wind speeds, with ~25% of towers downed by ~80mph winds in New York during Hurricane Sandy (Woetzel et al., 2020). Additional information can be found in a report published by the Public Safety and Homeland Security/ Bureau Federal Communications Commission (2018) which provides data about the impact of hurricane Harvey, Irma, Maria, and Nate on communication infrastructures.

Aside from purely physical damage, unaffected cellular network assets can also be overloaded following a disaster due to heavy traffic congestion, preventing the proper functioning of mobile voice and data services. Empirical evidence from wireless network operators demonstrates how wireless traffic sees a statistically measurable increase during extreme events (Jakubek, 2015), often exceeding the load each asset was provisioned for at the design stage, potentially leading to interrupted or severely degraded service.

Cellular network planning traditionally does not place much emphasis on potential natural disaster scenarios. A standard planning approach consists of three stages (Oughton et al., 2022a) where a network operator first undertakes high-level regional assessment of new asset deployments. Secondly, cluster-level evaluation is undertaken for a set of cells within a defined geographic proximity (e.g., 20 km²). Finally, high-fidelity link-level radio simulations are utilized to optimize cell site location, antenna designs, power levels, tower heights, pathloss, predicted interference and link margins. The aim here is to maximize coverage and capacity, while minimizing the cost of any necessary investment. As new spatial patterns of traffic demand emerge, the network operator iteratively redesigns the system of cells, by partitioning existing coverage areas and building new assets (Eiselt and Marianov, 2012). Due to cost considerations, moving existing major cell sites is generally avoided.

In terms of risk assessment, there has only been very limited research published in the literature focusing on deductive appraisals of future risks. One example utilizes event-based storylines to evaluate the impact of future climatic and socioeconomic conditions critical infrastructure in coastal flood areas (Koks et al., 2022). Critical infrastructure asset locations have been derived from OpenStreetMap, consisting of only 141,478 communication towers and 80,750 masts globally (Nirandjan et al., 2022).

There is also evidence suggesting fiber optic network assets can be vulnerable to natural hazards. Both surface and sub-sea water movements can severely damage the functionality of these assets. For example, cable fault data from Typhoon Morakot (2009) indicates that a major river flood, formed during two destructive sediment flows, led to the breaking of fiber cables in the Gaoping Canyon/Manila Trench, off the coast of Taiwan (Carter et al., 2012).

This literature review identified the (i) why mobile telecommunication infrastructure plays such an important role during disaster events, (ii) the different types of damage states that might occur to mobile cell tower sites, and finally (iii) identified some of the more recent deductive evaluations of this critical infrastructure sector. Now this review is complete, a method capable of answering the research questions will be outlined.

## 3. Method

A geospatial assessment method is developed to undertake this appraisal, with the parts outlined graphically in Figure 1. The approach consists of multiple steps, including the selection of scenario parameters, global hazard models and affiliated data, and then intersection processing of geolocated mobile cells to undertake a vulnerability assessment using exposure-damage curves to produce the required results. The method steps will now be articulated in detail.

Figure 1 Overview of modeling method

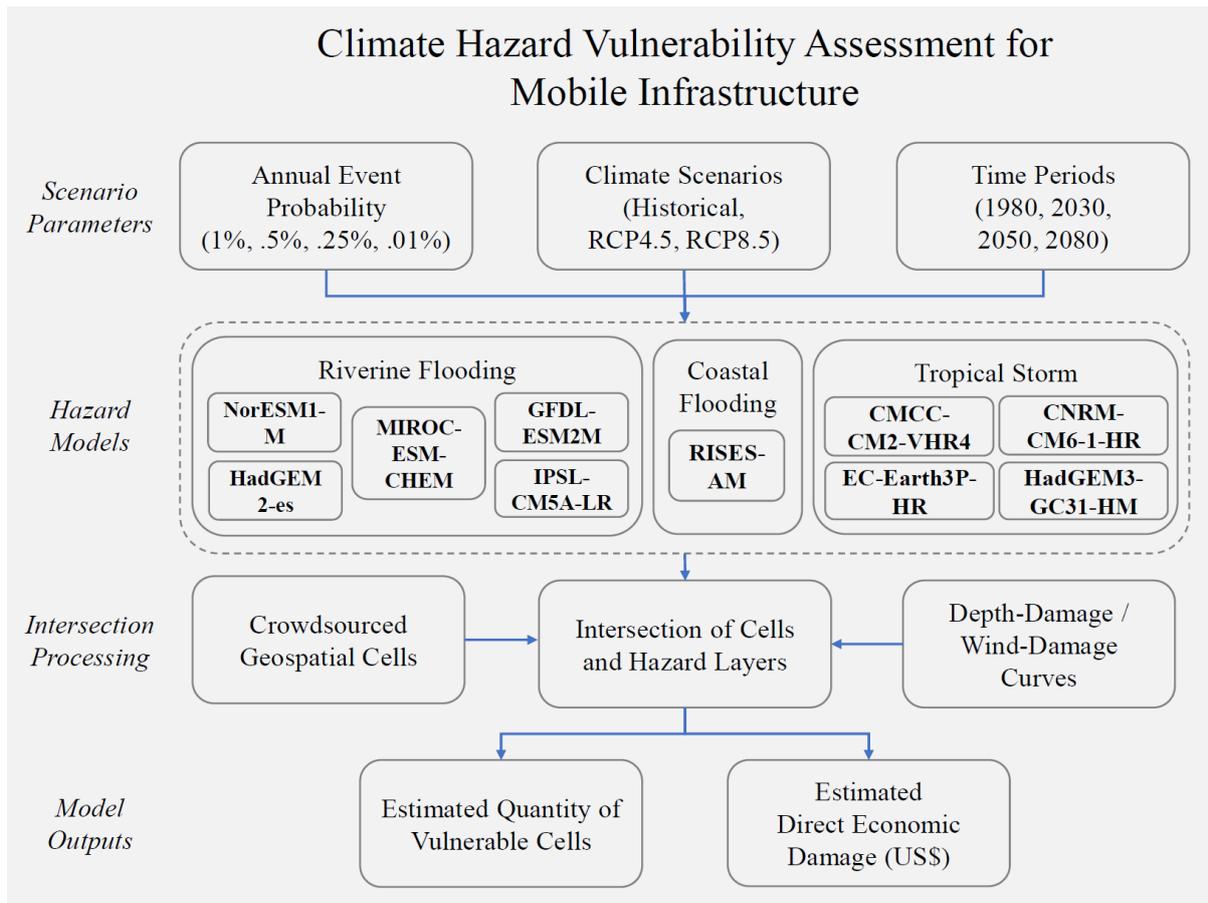

## 3.1. Climate hazard data

Global Aqueduct geospatial hazard datasets are obtained, selected for key scenario parameters of interest, and then processed for both riverine and coastal flooding layers (World Resources Institute, 2022). Data layers are selected for two different climate scenarios (as well as a 1980 historical baseline), based on the IPCC 5[th] assessment report. These include (i) Representative Concentration Pathway 4.5 (steady carbon emissions) (intermediate climate outcomes) (RCP 4.5) and (ii) Representative Concentration Pathway 8.5 (rising carbon emissions) (limited climate outcomes) (RCP 8.5).

The former scenario (RCP 4.5) is seen to be an optimistic future, reflecting considerable carbon mitigation by 2040 onwards. Whereas the latter (RCP 8.5) is essentially a pessimistic business-as-usual case (rising carbon emissions) with little carbon mitigation taking place (thus, Earth's current global trajectory is closest to RCP 8.5). Four main return periods are also selected for assessment, including events equating to 1% annual probability (1-in-100-year), 0.4% annual probability (1-in-250-year), 0.2% annual probability (1-in-500-year), and the most extreme 0.01% annual probability (1-in-1000-year).

In this analysis, coastal hazards represent storm surge flooding along coastlines with simulation projections from sea level change derived from the RISES-AM project. Three different sea rise scenarios are selected based on the default 95[th] percentile, as well as the 5[th] and 50[th] RISES-AM percentiles. Additionally, scenario layers which include estimated subsistence impacts are included in the analysis.

Riverine hazards in this evaluation represent overflow river flooding events, most likely to occur in river basins from excessive water loads. There are five common models widely available from different institutions for riverine flooding, including (i) GFDL-ESM2M from the US Geophysical Fluid Dynamics Laboratory (NOAA), (ii) HadGEM2-es from the UK Met Office Hadley Centre, (iii) IPSL-CM5A-LR from the French Institut Pierre-Simon Laplace, (iv) MIROC-ESM-CHEM from the Japanese Atmosphere and Ocean Research Institute (The University of Tokyo), National Institute for Environmental Studies, and Japan Agency for Marine-Earth Science and Technology, and finally (v) NorESM1-M from the Bjerknes Centre for Climate Research, at the Norwegian Meteorological Institute. This analysis does not aim to preference a specific model, instead opting to use these capabilities as an ensemble to obtain broad statistical estimates given each of these five global circulation model options.

Figure 2 and Figure 3 illustrate the sum of flooded pixels by continent and flood region, respectively, by annual probability, year, and climate scenario. The mean average is taken across the different riverine models.

Figure 2 illustrates large increases in coastal flooding over the different climate scenarios, with a clear upward trend in higher emissions environments. In comparison, riverine flooding sees a more mixed result, with either only minor increases in flooded area (e.g., North America or Sub-Saharan Africa), or a static trend (e.g., Western Europe). The literature has already identified highly heterogenous riverine flooding impacts (especially for Europe) (Blöschl et al., 2019), with high variability between flooding models leading to greater uncertainty in flooding impacts (especially for high emissions scenarios) (Dottori et al., 2018).

Return period maps for maximum wind speeds experienced during tropical cyclones are obtained from Bloemendaal et al 2020 and 2022 for present and future (2050s, SSP5-8.5) climate conditions, for a fixed set of return periods between 10 and 10,000 years, at 10 km resolution, for each cyclone basin (Bloemendaal et al., 2022, 2020). The future climate conditions are derived from a set of general circulation models: CMCC-CM2-VHR4, CNRM-CM6-1-HR, EC-Earth3P-HR and HadGEM3-GC31-HM. These models are evaluated and briefly described in the literature (Roberts et al., 2020). Each set of return period maps is calculated from 10,000 years of synthetic tracks modeled using the Synthetic

Tropical cyclOne geneRation Model (STORM). The global maximum windspeed return period maps are mosaiced together across all basins to provide global GeoTIFFs (Russell, 2022).

*Figure 2 Coastal and riverine flooding impacts by continent*

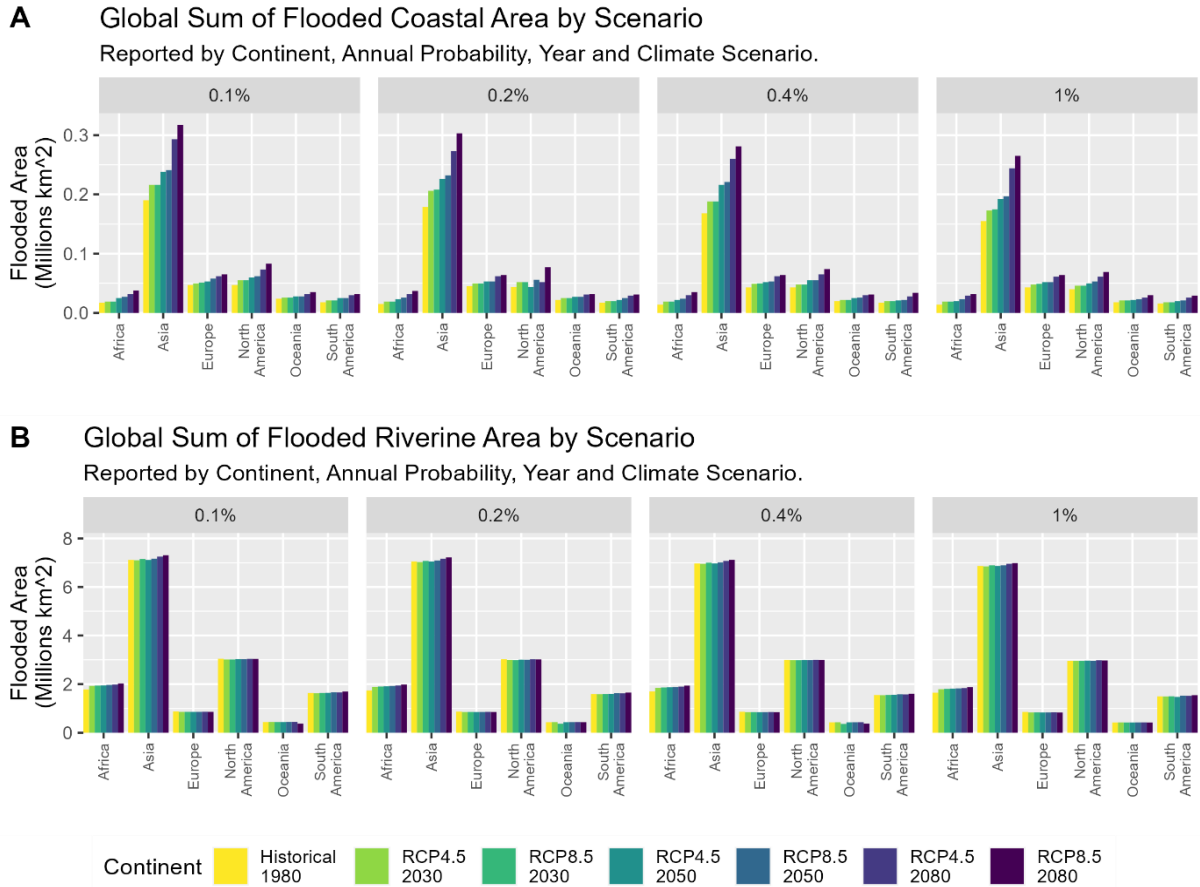

## 3.2. Mobile infrastructure asset data

Crowdsourced data for December 2022 are gathered from OpenCelliD for current mobile infrastructure networks, covering 7.6 million cellular assets for 2G, 3G, 4G and 5G. The Global System for Mobile Communications (GSM) is regarded more commonly as the second cellular generation or '2G'. following standardization in 1991, commercial deployments were launched across the world in the 1990s onwards, with the technology offering revolutionary voice and text Short Messaging System (SMS) services. A total of 10.3% of the cells in the dataset as 2G, equating to a total of 0.78 million. Although outdated in many high-income countries, 2G is still very much widely used technology, especially in low- and middle-income countries.

The third cellular generation known as '3G' consists of multiple technologies, with the main standard known as the Universal Mobile Telecommunications System (UMTS). Standardized in 2001, commercial launches widely took place throughout the 2000s. The key benefit of 3G deployment was the introduction of basic data capabilities, enabling users to access a peak connection speed of ~3 Mbps. This enables web browsing and very basic low-resolution video functions. Approximately 41.9% of the assets in the dataset are 3G, equating to 3.18 million cells.

*Figure 3 Coastal and riverine flooding by flood region*

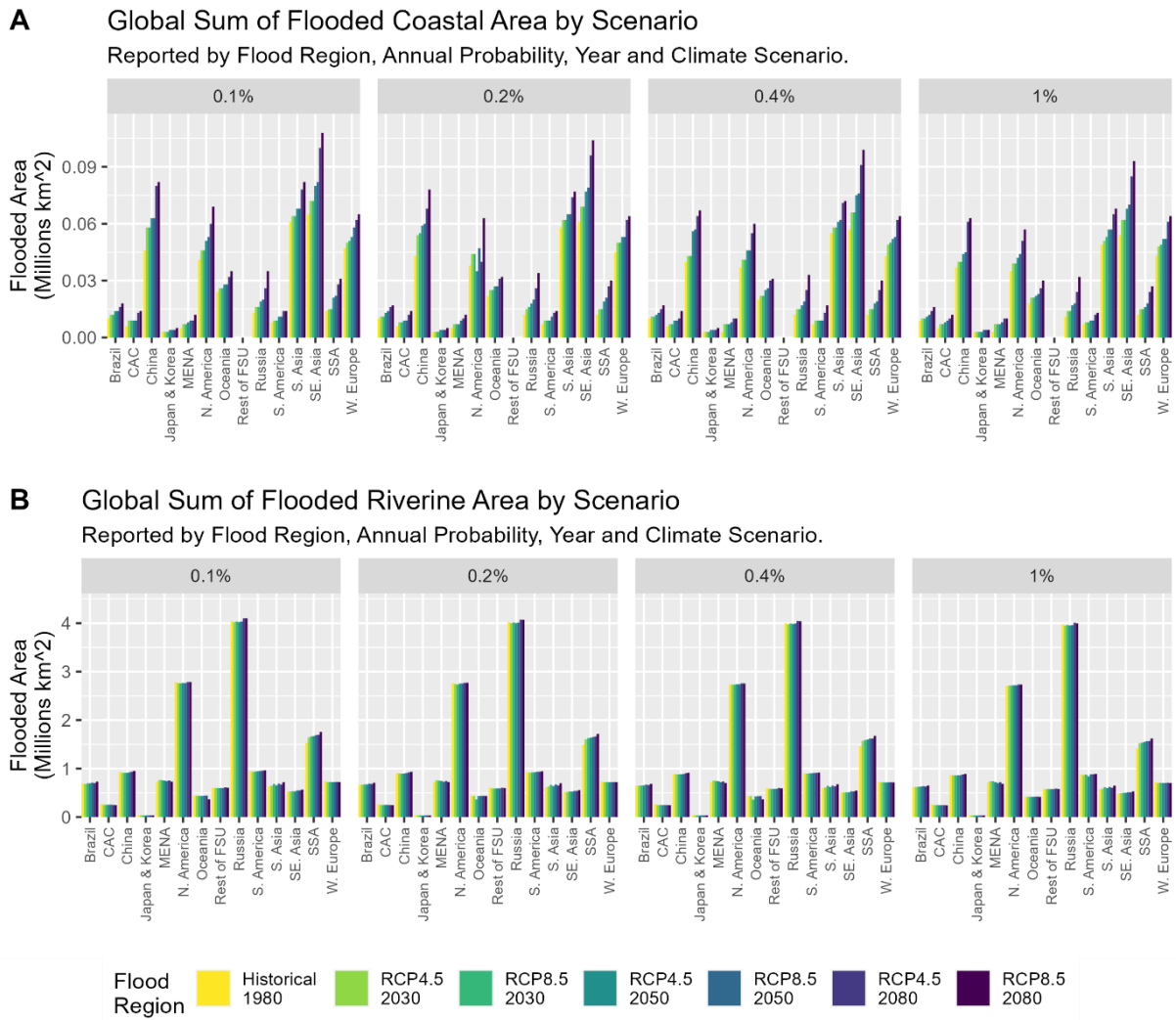

The fourth generation of cellular technology ('4G') was standardized in 2009 with the name Long Term Evolution (LTE) and brought mass market mobile broadband to billions of consumers around the world (mainly deployed in the 2010s onwards). This was driven by the introduction of smartphones as a key general-purpose technology, largely resulting from Apple's pioneering iPhone. With peak speeds capable of up to ~100 Mbps, users have the capability to download high-definition video while mobile, as well as carry out a large range of online activities, most notably sharing and consuming high-quality content across a range of social media platforms. The dataset contains 3.63 million 4G LTE assets, comprising roughly 47.8%.

Finally, the introduction of the fifth generation of cellular technology known as 5G has only fully taken place in recent years, from approximately 2020 onwards (Oughton et al., 2021). The standardized radio technology is referred to as the 5G New Radio (NR) standard (Cave, 2018; da Ponte et al., 2023; Lehr et al., 2021). Three key use cases exist for users, including enhanced mobile broadband, ultra-reliable low-latency communications, and massive machine type communications (Bauer and Bohlin, 2022; Oughton and Lehr, 2022). Given the limited number of countries with commercial 5G launches, less than 0.04% of the sample consists of 5G NR cells (<2,718 assets). However, over the next decade, older cells will be decommissioned and replaced with 5G NR equipment. Therefore, this number is expected to significantly increase in coming years, especially as higher-income countries turn off older generations such as 2G GSM.

### 3.3. Vulnerability analysis

Mobile cellular towers generally consist of the three main design types identified within the literature review, including monopoles, self-supported structures, or guyed structures. These assets usually rely on steel or aluminum for the structural frame, utilizing concrete at the base of the tower either for the foundation plinth or to secure guy lines. Vulnerability to hazards and the consequential damage cost is estimated using a damage curve approach, which is a standard way in the literature to relate, for example, the depth of flooding inundation or wind speed to a potential direct damage cost (DHS and FEMA, 2011; Miyamoto International and World Bank, 2019; Movahednia et al., 2022; Sánchez-Muñoz et al., 2020). However, there are not readily available damage curves for communication towers. Therefore, existing studies in the literature adopt a similar approach to assessing the vulnerability of electricity infrastructure assets, such as distribution poles and transmission pylons which share similar structural characteristics (Kok et al., 2005). Indeed, some studies utilize curves from the US HAZUS model (DHS and FEMA, 2011) for electricity infrastructure, to assess the vulnerability of communications infrastructure assets (Koks et al., 2022), as will be adopted here. Unfortunately, within the US HAZUS model, the development of fragility relationships specifically for telecommunications infrastructure assets based on inundation has been stated as "deferred to a later date" (p7-10), with no current update on the status of these improvements (DHS and FEMA, 2011) (thus, identifying an important area of future research).

Booker et al (2010) define fragility curves for each failure mode of cellular sites due to wind speed. These fragility curves are at the level of individual failure modes (e.g., foundation failure) rather than at the level of a complete cellular tower site because, different failure modes have different implications for the functionality of the cellular site. For example, loss of cellular antenna at a given tower would prevent that tower from receiving or broadcasting calls, but it would still be able to relay calls from other towers if its microwave dish or fiber–optic link was functional. However, if the tower experiences structural failures, it would not be able to service incoming calls, outgoing calls, or relay calls through its microwave dish, but it would still be able to serve as a relay through its fiber–optic cable if the necessary power supply and supporting equipment were functional (Booker et al., 2010).

The investment costs for rebuilding mobile cells are adapted from the existing literature (Oughton et al., 2022b), to enable damage costs for different sized events and climate scenarios to be estimated. Broadly, the cost of building a new three-sector 4G macro cell site is approximately $100,000, equating to approximately $33,333 total capital expenditure per cell. Thus, a cell with a flooding depth of 0.6 meters and a damage quantity of 0.5, leads to a direct damage cost estimate of $16,666.5. As identified in the literature review, the specifics of the damage depend on the cell tower construction. At the lower end of the fragility curve, damage to electronic equipment could take place. Whereas in more severe outcomes, this could include the full destruction of both active electronic radio equipment and passive civil engineering tower structures, requiring a total rebuild of a site.

The intersection process is carried out practically using a range of python packages. For example, managing spatial point data is carried out via *geopandas*, and then raster-based queries of the hazard layers are undertaken using a combination of *rasterio* and *rasterstats*. For each lower layer region, this information is exported to a comma-separated value file, and then later aggregated for reference in the following results section.

## 4.     Results

The results will now be reported for coastal flooding, riverine flooding, and tropical cyclone hazards. First a set of metrics is presented which summarizes the main hazard differences per scenario. Next, the estimated results for mobile broadband infrastructure vulnerability are presented for each threat.

### 4.1.     Vulnerability to coastal flooding

The number of vulnerable cells at risk from coastal flooding is reported in Figure 4 (A) by year, and annual probability for various climate scenarios. For coastal flooding events with a 1% annual probability, equating to a 1-in-100-year event magnitude, 52.2 thousand assets are at risk when considering the historical 1980 climate scenario. However, by 2080 the number of vulnerable assets is estimated to increase to 79.9 thousand in the RCP4.5 scenario, and 87.8 thousand in the RCP8.5 scenario (an increase of 53% and 68%, respectively). In contrast, a 1-in-1000-year event with a 0.1% annual probability is estimated to affect 64.5 thousand annual cells in the 1980 historical climate baseline. In the RCP4.5 scenario by 2080, this leads to an estimated increase to 99.7 thousand vulnerable cells (a 55% increase), with up to 109.9 thousand cells vulnerable in the RCP8.5 climate scenario (a 70% increase).

In Figure 4 (B), the analytical results are presented for the direct economic damage caused to assets from coastal flooding. The historical baseline for an event with a 1% annual probability (a 1-in-100-year event) is estimated at $1.18 billion, with this figure increasing to $1.83 billion by 2080 in the RCP4.5 scenario, and $2.07 billion in the RCP8.5 scenario (an increase of 55% and 75%, respectively). For larger events, such as those with a 0.1% annual probability (a 1-in-1000-year event) the historical baseline estimates a $1.51 billion impact, with this rising by 2080 to $2.41 billion in the RCP4.5 scenario (an increase of 60%), and $2.69 billion in the RCP8.5 scenario (an increase of 78%).

*Figure 4 Mobile infrastructure vulnerable to coastal flooding*

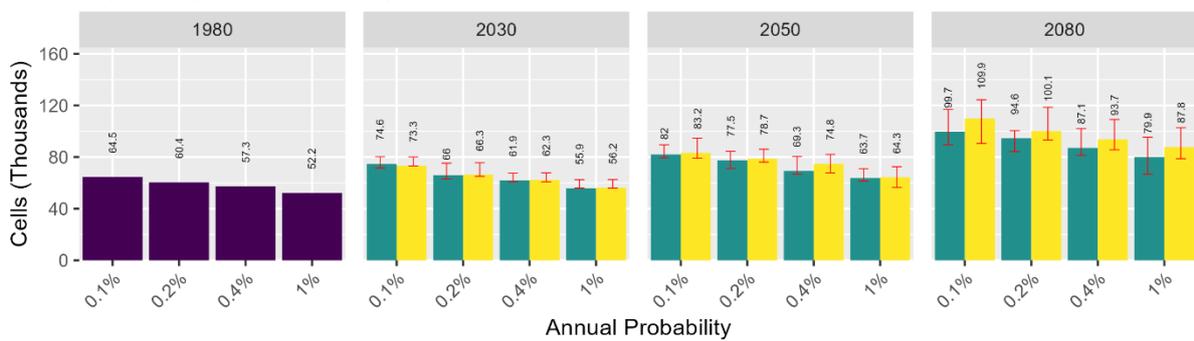

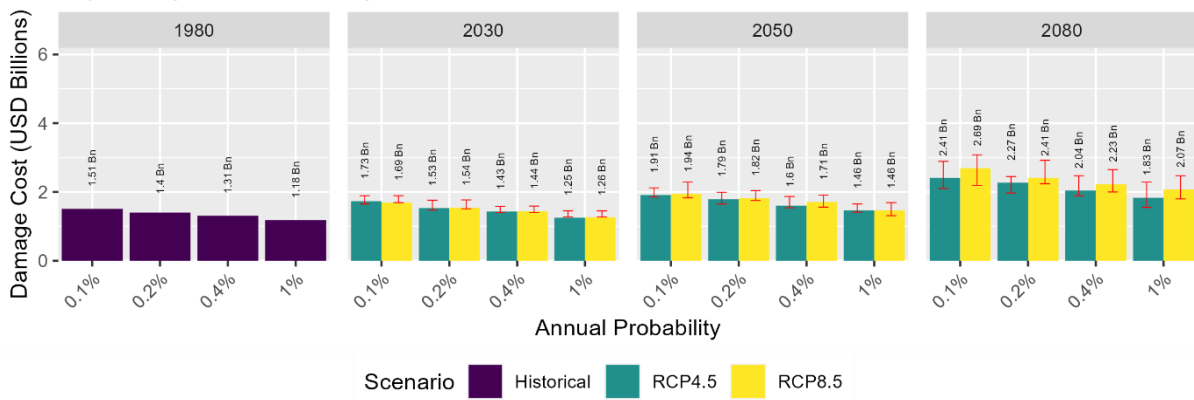

To provide greater geospatial insight, Figure 5 illustrates the estimated coastal flooding impacts for the RISES-AM model. For the RCP8.5 scenario in 2080, an event with a 0.01% annual probability (1-in-1000-year), large impacts are visible in Western Europe (e.g., The Netherlands), East Asia (e.g., China and Japan), Southeast Asia (e.g., Indonesia), North Africa (e.g., Egypt), and Latin America (e.g., Brazil) (>$1 million per local statistical area). The highest quantity of estimated vulnerable cells and affiliated damages are in Asia (64 thousand equating to 58%), Europe (35 thousand equating to 32%), and North America (9 thousand equating to 8%).

As the underlying crowd-sourced data contains more users and thus more asset data for higher income countries, the results see the largest impacts overall in this income group. For example, across the scenarios approximately 53% of the vulnerable cells assessed for coastal flooding are in high income countries, compared to an estimated 35% upper middle-income countries, and 12% in lower middle-income countries. The lack of cell data available for low-income countries leads to very low estimated impacts for this income group (<1%).

*Figure 5 Estimated coastal flooding damage cost*

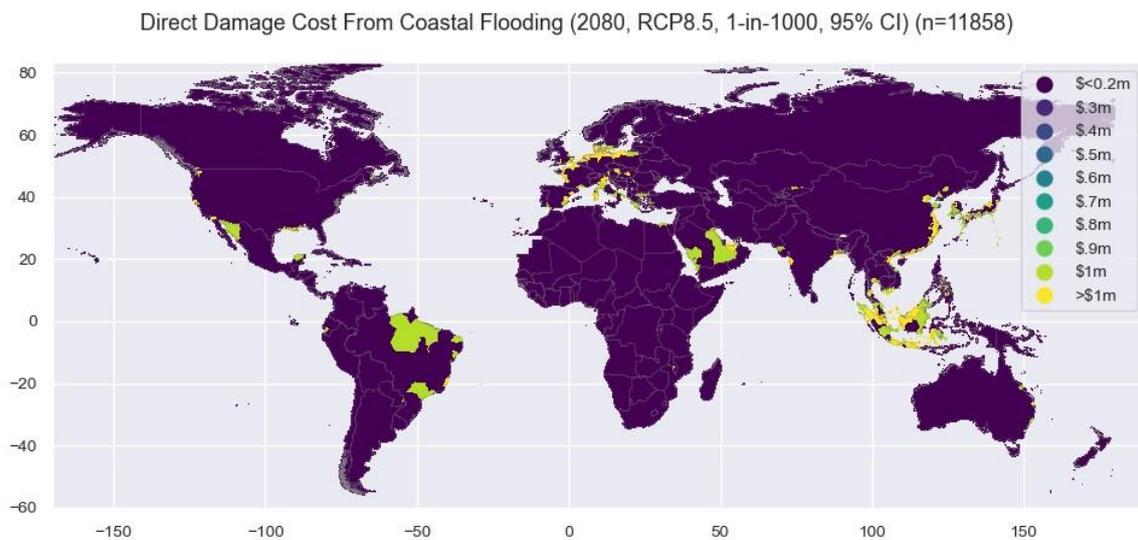

## 4.2. Vulnerability to riverine flooding

Next, Figure 6 (A) reports the number of cells vulnerable to riverine flooding. For this hazard there is less of a pronounced trend between more severe climate scenarios and increasing cellular asset vulnerability. For example, the historical baseline from 1980 estimates a total number of vulnerable cells to be 0.99 million for events with a 1% annual probability (a 1-in-100-year return period), up to 1.05 million for events with a 0.1% annual probability (a 1-in-1000-year return period). Thus, vulnerability estimates are estimated to see mixed impacts over time. Approximately, 0.99 million cells at risk in 2080 in the RCP4.5 scenario, and 1 million cells at risk in 2080 in the RCP8.5 scenario (a 1% increase), for events with 1% annual probability. Similar results occur for the 1-in-1000-year return period, equating to 0.1% annual probability, with 1.05 million cells at risk in 2080 in the RCP4.5 scenario, and 1.06 million cells at risk in 2080 in the RCP8.5 scenario.

Therefore, in Figure 6 (B) the direct economic damages follow this decreasing trend. For example, in the less severe case of events with a 1% annual probability (a 1-in-100-year return period), the estimated damage ranges from $14.3 billion in the historical 1980 baseline to 2080 estimates of $14.3 billion in the RCP4.5 climate scenario (no change), and $14.7 billion in the RCP8.5 climate scenario (a 3% increase). In the more severe event case of 0.1% annual probability (a 1-in-1000-year return period),

the estimated damage ranges from $18 billion in 1980, to 2080 estimates of $17.8 billion for the RCP4.5 climate scenario (a 1% decrease) and $18.3 billion for the RCP 8.5 climate scenario (a 2% increase).

*Figure 6 Mobile infrastructure vulnerability to riverine flooding*

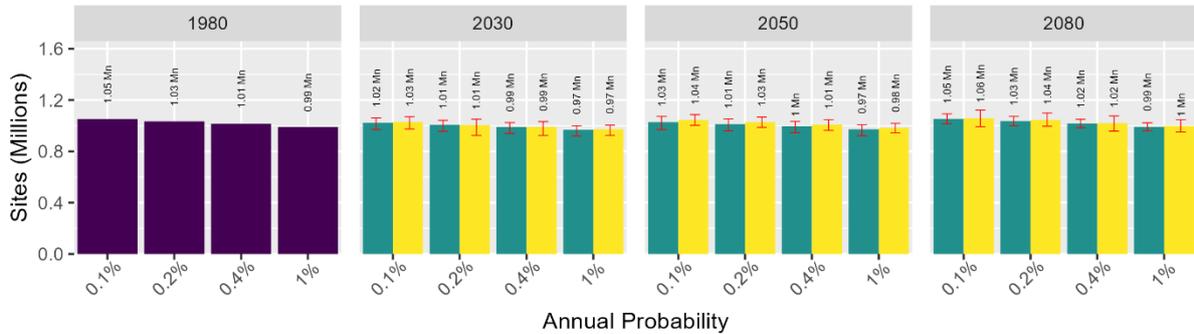

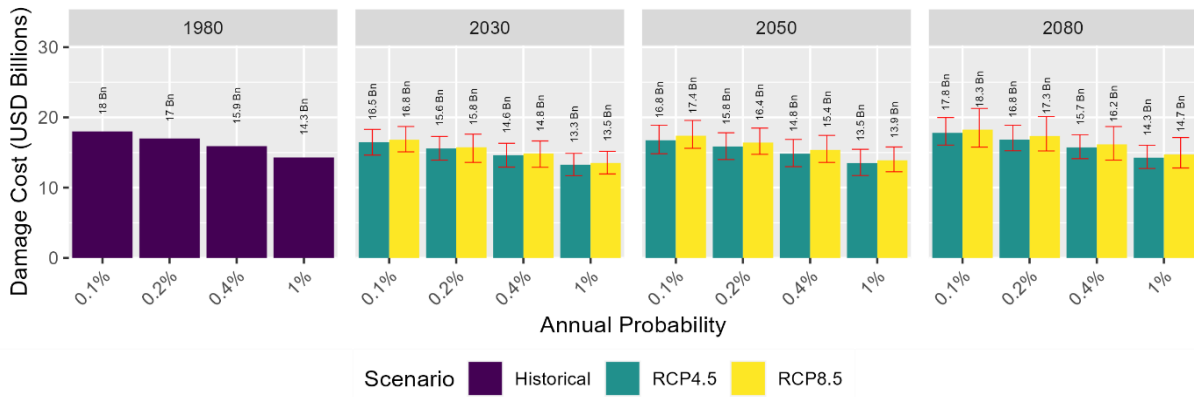

Riverine impacts are visualized in Figure 7 for the RCP8.5 scenario in 2080, with an annual probability of 0.01% (1-in-1000-year). The estimated damage values represent the mean average across the five model variants. Large damage impacts are estimated for Eastern Europe (e.g., Germany, Poland etc.), Latin America (e.g., Brazil), southeast Asia (e.g., Indonesia), east Asia (e.g., Japan), and central Asia (e.g., Russia, Kazakhstan) (>$1 million per local statistical area).

Again, the largest impacts are present in high income countries (approximately 59%), followed by 29% in UMCs, 12% in LMCs and <1% in LICs. Compared to coastal flooding, the level of vulnerability to riverine flooding is more evenly spread across regions. For example, Asia comprises approximately 42% of the impacts, followed by 28% in Europe, 22% in North America, 4% in South America, and 2% each in Africa and Oceania.

*Figure 7 Estimated riverine flooding damage cost*

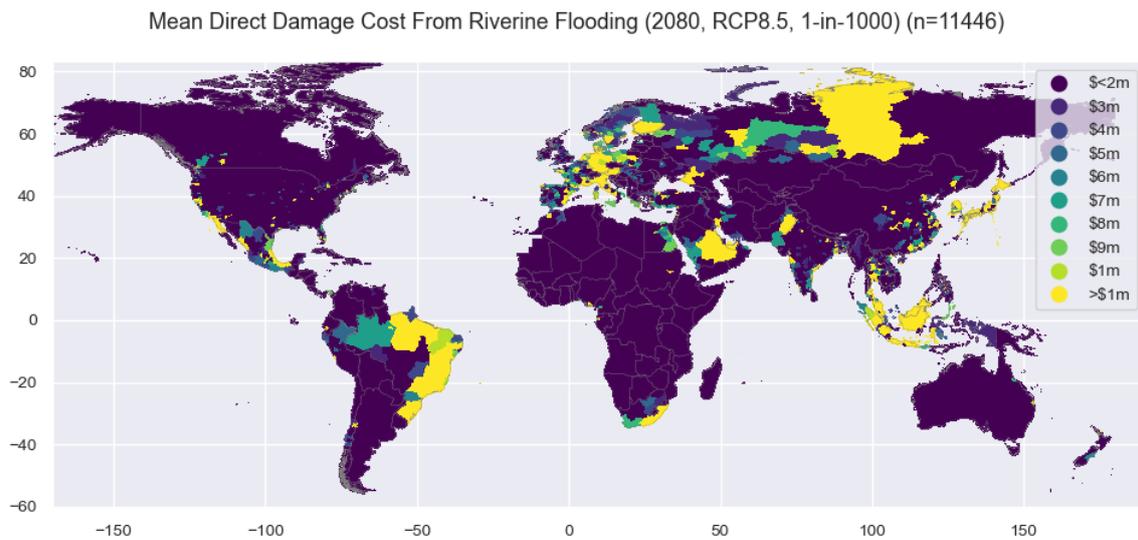

## 4.3. Vulnerability to tropical cyclones

Figure 8 (A) illustrates the mobile cell vulnerability results for tropical cyclone hazards, for different years and annual probabilities of occurrence. Two scenarios exist, including a 1980 historical climate scenario, and then a future 2050 climate scenario equivalent to RCP8.5. Approximately 0.21 million cells are estimated to be affected in the 1980 historical baseline for an event with a 1% annual probability, raising to 0.73 million in 2050 for an RCP8.5 scenario (an increase of 248%). This compares to 1.98 million cells in the 1980 historical baseline for an event with a 0.01% annual probability, increasing to 2.26 million by 2050 for the RCP8.5 scenario (an increase of 14%).

The direct damage cost is also illustrated for a set of historical scenarios, as reported in Figure 8 (B). The results indicate very large differences in vulnerability. For example, for the smallest event size, with a 1% annual probability (a 1-in-100-year return period), the total direct economic damage is estimated at $0.03 billion for the 1980 historical scenario, with 56% of this impact in Asia and 43% in North America. In contrast, under the more severe 2050 climate scenario, the estimated damage reaches $0.09 billion (an increase of 200%) (with a model range of $0.08-0.1 billion), with 83% of the impact in Asia, and 16% in North America. In contrast, in the largest event size with a 0.01% annual occurrence (a return period of 1-in-1000-years), the historical vulnerability is estimated at $0.7 billion, compared to the future 2050 climate scenario estimate of $1.01 billion (a 44% increase) (with a model range of $0.82-1.17 billion). Approximately 32% of the 2050 damage estimated is in Asia, with 38% in North America, 20% in Europe, and 6% in South America.

*Figure 8 Mobile infrastructure vulnerability to tropical cyclone hazards*

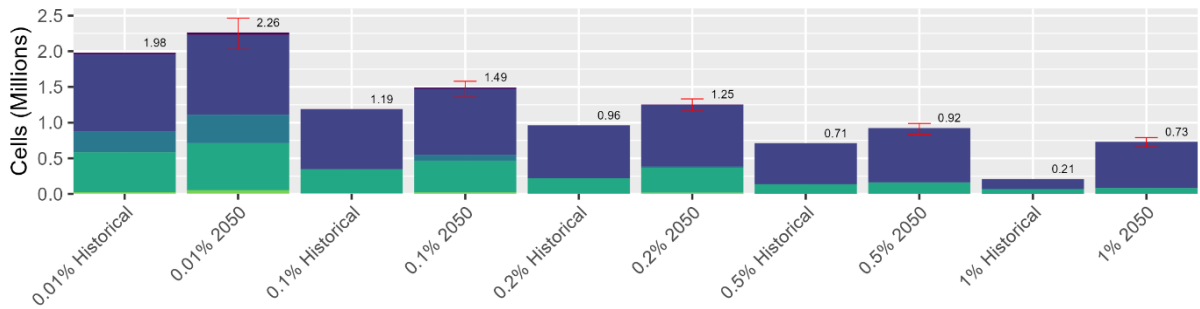

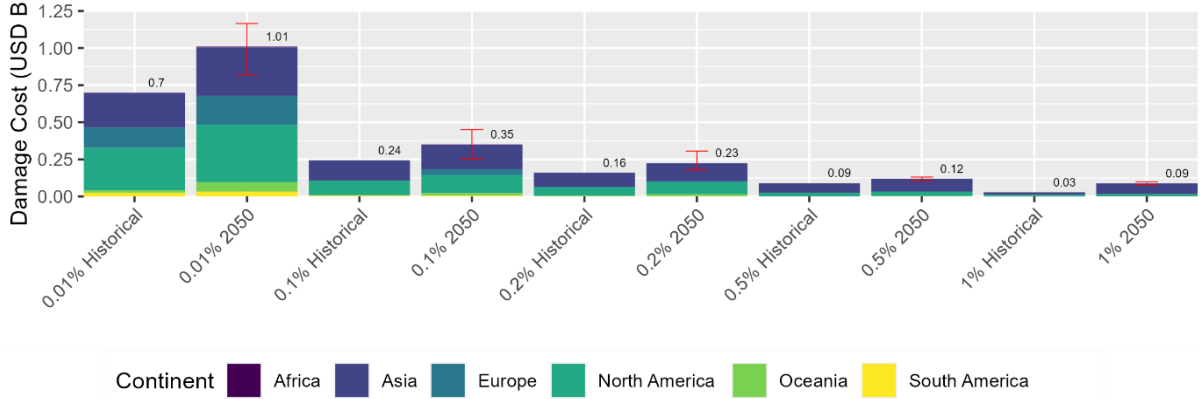

In Figure 9 the estimated mean damage cost is visualized globally for an event with a 0.01% annual probability (1-in-1000-year), for the latest time point available (2025), in an RCP8.5 scenario. The largest damage impacts are visible in East Asia (e.g., Japan, South Korea etc.), South Asia (e.g., India, Bangladesh, Pakistan etc.), Southern Europe (e.g., Portugal, Spain etc.), North Africa (e.g., Morocco), and in North and Central America (e.g., the USA, Mexico, Dominican Republic, Venezuela etc.). While Japan and South Korea indicate high levels of potential damage nationally, in the Americas, key vulnerable regions include Florida in the USA, and Yucatan and Tamaulipas in Mexico. Equally, the Andalucía region of southern Spain (containing Seville) sees elevate vulnerability, along with the Sindh area of Pakistan (containing Karachi).

*Figure 9 Estimated tropical cyclone damage cost*

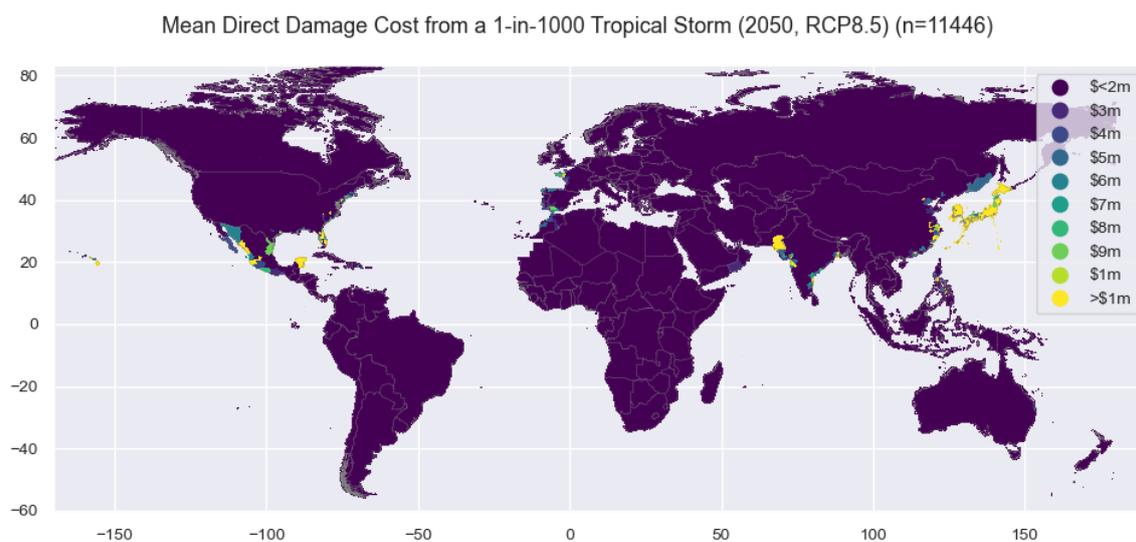

## 5.  Discussion

Having presented the estimated results for the three main climate hazards evaluated, this section undertakes a discussion of the key findings in relation to each research question. Each one of the research questions will be returned to sequentially. For the 7.6 million cellular assets assessed for 2G, 3G, 4G and 5G, the results vary strongly for countries, income groups and regions. These estimates reflect both the spatial hazard differences across countries, and the spatial distribution of the cells present in the underlying data.

*How many cellular assets are potentially vulnerable to climate hazards, such as coastal flooding, riverine flooding, and tropical cyclones?*

For coastal flooding, the quantity of cells vulnerable from an event with a 1% annual probability is estimated at 52.2 thousand for the historical 1980 climate scenario, increasing by 2080 to 79.9 thousand for RCP4.5 (a 53% increase) and 87.8 thousand for RCP8.5 (a 68% increase). Moreover, an event with a 0.1% annual probability in the 1980 historical climate baseline is estimated to affect 64.5 thousand cells, versus 99.7 thousand in the RCP4.5 scenario (a 55% increase), and 109.9 thousand cells in the RCP8.5 climate scenario (a 70% increase).

For riverine flooding, estimates suggest that for a 1980 historical baseline the total number of vulnerable cells for an event with a 1% annual probability is 0.99 million. Under the RCP4.5 scenario this remains static at 0.99 million, whereas there is a marginal increase for the RCP8.5 scenario to 1 million vulnerable cells (a 1% increase). For events with a 0.1% annual probability, approximately 1.05 million cells are at risk for the 1980 historical baseline, remaining static at 1.05 million cells in the RCP4.5 scenario in 2080, and increasing to 1.06 million cells in the RCP4.5 scenario (a 1% increase).

The estimates reached for tropical cyclones suggest many cells could be affected. For example, for an event with a 1% annual probability, 0.21 million affected cells are estimated as experiencing some form of damage, with this increasing by 2050 in an RCP8.5 scenario to 0.73 million (a 248% increase). For a much larger event with a 0.01% annual probability, the 1980 historical baseline sees 1.98 million cells potentially affected, increasing by 2050 for the RCP8.5 scenario to 2.26 million (a 14% increase).

*What magnitude of economic damage may occur from different climate hazard scenarios, for example from coastal flooding, riverine flooding, and tropical cyclones?*

The direct economic damage estimates for coastal flooding range from $1.18 billion an event with a 1% annual probability in the 1980 historical baseline, up to $1.83 billion by 2080 in the RCP4.5 scenario, or $2.07 billion in the RCP8.5 scenario (an increase of 55% and 75%, respectively). Compared to a more extreme scenario with a 0.1% annual probability, a $1.51 billion estimate is produced for the 1980 historical baseline, rising to $2.41 billion in the RCP4.5 scenario (an increase of 60%), and 2.69 billion in the RCP8.5 scenario (an increase of 78%) by 2080.

Riverine flooding produced a different trend, as the less severe case with a 1% annual probability had a direct damage estimate of $14.3 billion in the historical 1980 baseline, compared to $14.3 billion in the RCP4.5 climate scenario, and $14.7 billion in the RCP8.5 climate scenario in 2080. In the more severe event case of 0.1% annual probability (a 1-in-1000-year return period), the direct damage estimate ranges from $18 billion in the 1980 historical baseline, to $17.8 billion for the RCP4.5 climate scenario and $18.3 billion for the RCP 8.5 climate scenario in 2080.

Finally, for tropical cyclone event sizes with a 1% annual probability the direct damage estimate ranges from $0.03 billion in the 1980 historical baseline, to $0.09 billion in 2050 under an RCP8.5 scenario (with a model range of $0.08-0.1 billion), increasing by 200%. For a much larger event with a 0.01% annual occurrence, the direct damage estimate for the 1980 historical baseline is $0.7 billion, rising significantly to $1.01 billion (with a model range of $0.82-1.17 billion) for a future 2050 climate scenario, increasing by 44%.

## 6.  Conclusions

In this paper a global assessment was undertaken evaluating the potential impacts to mobile cellular infrastructure for a variety of climate scenarios. Using open crowdsourced data equating, to 7.6 million 2G, 3G, 4G and 5G cellular assets, a variety of climate scenarios were explored, quantifying potentially damaged assets and direct damage costs. The results were reported for a range of event sizes with different annual probabilities. The paper contributes to the literature one of the first global assessments of mobile cellular infrastructure for climate related hazards.

In the baseline historical scenarios for an event with a 0.1% annual probability, the largest number of affected cells was found to be from tropical cyclones (1.98 m), followed by riverine flooding (1.05 m) and coastal flooding (64.5 thousand). However, the number of affected cells did not necessarily yield the same magnitude in terms of direct damage impacts. For example, the historical estimated damage cost was $0.7 billion for tropical cyclones, $18 billion riverine flooding, followed by $1.51 billion for coastal flooding. The results suggest that many cells can be affected by high winds from tropical cyclones but may only receive very minor damage.

When examining the impacts under future climate scenarios, the results suggest that tropical cyclones and coastal flooding see the largest increases in the most severe high emissions futures explored. For example, for a 0.01% annual probability event under RCP8.5, the number of affected cells is estimated to rise to 2.26 million for tropical cyclones (an increase of 14%), equating to $1.01 billion in direct damage (an increase of 44%). Equally, for coastal flooding the number of potentially affected cells for an event with a 0.01% annual probability under RCP8.5 is 109.9 thousand (an increase of 70%), equating to direct damage costs of $2.69 billion (an increase of 78%). Riverine flooding saw estimates remain relatively static, due to many mountainous areas having lower winter snowfall, and therefore smaller meltwater runoff leading to flooding events.

The findings demonstrate the need for risk analysts to include mobile communications (and telecommunications more broadly) in future critical national infrastructure assessments. Indeed, this paper contributes a proven assessment methodology to the literature for use in future research for assessing this critical infrastructure sector.

## Acknowledgements

This research endeavor has been supported by funding from multiple sources. These include the World Bank Digital Development Global Practice (as background evidence for the report Catalyzing the Green Digital Transformation in Low- and Middle-Income Economies, and Climate Change Development Reports for Azerbaijan, Kenya, and the Democratic Republic of Congo), and the Insurance Development Forum (as part of the Global Resilience Index Initiative).

Supplementary information

S1 Hazard scenario metrics

In S1 (A) the mean inundation depth is presented for coastal flooding, for four different annual probabilities, disaggregated by continent. Generally, the historical mean depth for 1980 is lower than the forecast years, across the two climate scenarios. For example, for a 1-in-100-year event size with a 1% annual probability, the 1980 mean depth is 4.58 m based across all flooded 1 km² tiles. In contrast, this value increases to 5.44 m by 2080 for the RCP4.5 scenario, or 5.66 m in the RCP8.5 scenario, again for the final year of the assessment period. In contrast, a larger event equating to a 1-in-1000-year return period, with a 0.1% annual probability, the historical mean depth is 4.77 m, increasing by 2080 to 5.87 m in RCP4.5, and 6.09 m in RCP8.5. The hazard dynamics illustrated are consistent with *a priori* expectations for how climate change will affect coastal flooding.

*S1 Hazard variance by scenario*

**A**  Global Mean Coastal Flooding Depth by Scenario
Reported by Annual Probability, Year and Climate Scenario.

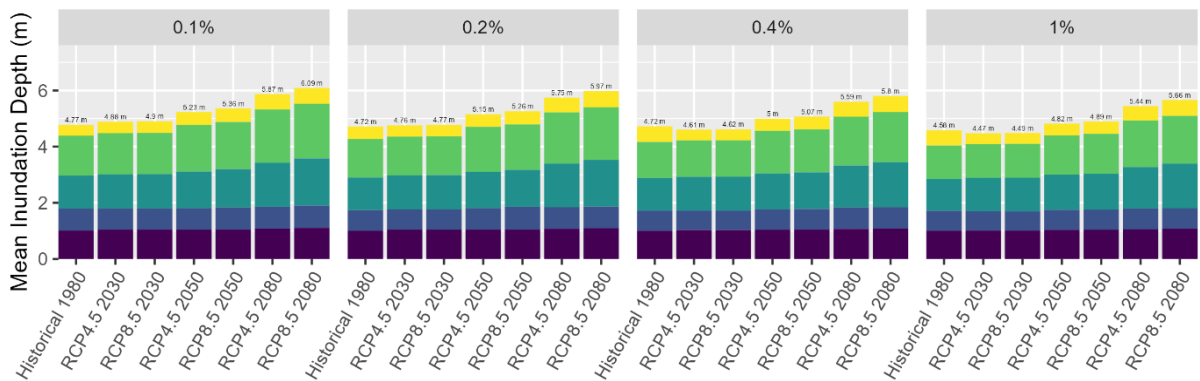

**B**  Global Mean Riverine Flooding Depth by Scenario
Reported by Annual Probability, Year and Climate Scenario.

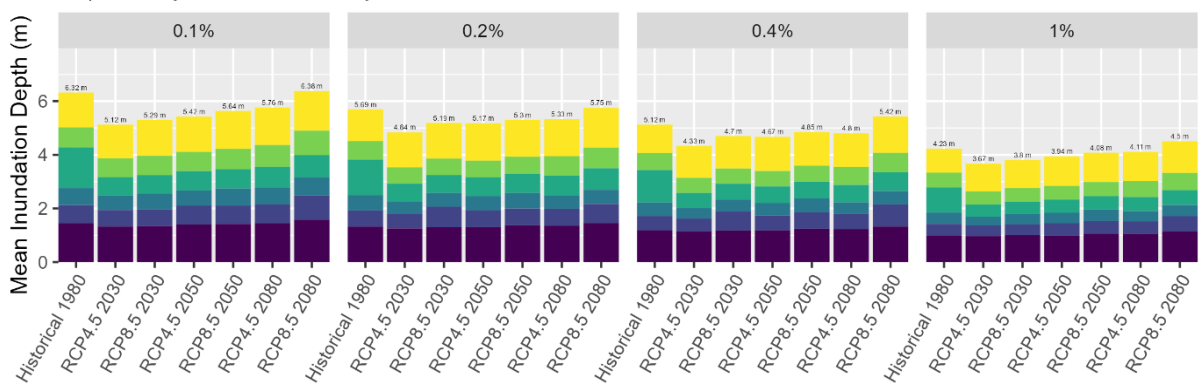

In S1 (B) the results are presented for riverine flooding by continent, focusing on the mean inundation depth for all flooded 1 km² tiles. In the 1980 historical baseline, for riverine flooding events with a 1% annual probability (1-in-100-year return period), a mean value of 4.23 m is obtained, and contrasts with 4.11 m in 2080 in the RCP4.5 scenario, and 4.5 m in the RCP8.5 scenario. These mean averages are logically higher for larger events with lower annual probabilities, such as a 1-in-1000-year magnitude (0.01% annual probability). For example, in the 1980 historical baseline the mean inundation depth is 6.32 m, rising to 5.76 m in the RCP4.5 scenario, and 6.36 m in the RCP8.5 scenario.

Whereas in the coastal mean averages, there is a clear relationship from a more severe climate scenario leading to increased flooding, this is not so clearly pronounced for riverine flooding. Yet, this is consistent with other studies in the literature which have demonstrated that some regions may see increased flooding, while others may see decreasing flooding levels. For example, a warmer climate may lead to lower winter snow levels in mountainous areas, consequently producing smaller meltwater volumes in spring and summer months in many catchment areas. Thus, the mean average inundation depth may be static or declining in such circumstances.